\documentclass[pre,twocolumn,preprintnumbers,amsmath,amssymb,nofootinbib,floatfix]{revtex4}

\makeatletter
\def\graphicscale{\twocolumn@sw{0.3}{0.4}}
\def\graphicthreescale{\twocolumn@sw{0.3}{0.4}}

\begin{document}
 \title{Off-equilibrium scaling driven by a time-dependent magnetic field in O$(N)$ vector models.}
  \author{Stefano Scopa$^{1,2}$ \\ email: stefano.scopa@univ-lorraine.fr}
 \address{$^1$Dipartimento di Fisica dell'Universit\'a di Pisa, Largo Pontecorvo 3, I-56127 Pisa, Italy}
 \address{$^2$ Institut Jean Lamour, Dpt. P2M, Groupe de Physique Statistique, Universite ́ de Lorraine, CNRS, B.P. 70239, F-54506 Vandoeuvre les Nancy Cedex, France}
 \date{\today}

 \begin{abstract}
 We investigate the off-equilibrium dynamics of a spin system with O$(N)$ symmetry in $2<d<4$ spatial dimensions arising by the presence of a slowly varying time-dependent magnetic field $h(t,t_s)\approx t/t_s$, $t_s$ is a time scale, at the critical temperature $T=T_c$ and below it $T<T_c$. After showing the general theory, we demonstrate the off-equilibrium scaling  and we formally compute the correlation functions in the limit of large $N$.  We derive the off-equilibrium scaling relations for the hysteresis loop area and for the magnetic work done by the system when the magnetic field $h(t,t_s)$ is varied across the phase transitions cyclically in time. We also investigate the first deviations from the equilibrium behavior in the correlation functions checking the consistence for an exponential approach. 
 \end{abstract}

 \maketitle
 \section{Introduction}
Phase transitions generally occur by varying the external fields across their critical values. But when statistical systems are driven through a critical point by time-dependent external fields, they show off-equilibrium behaviors. The emergences of these behaviors are related to the phenomenon of critical slowing down i.e. to the presence of large-scale modes which cannot adapt themselves to the changes of the external parameters, even when the time-scale $t_s$ of the variations of the external fields becomes very large, $t_s\rightarrow\infty$. The study of continuous phase transitions induced by slow variations of the external fields is generally called \textit{Kibble-Zurek} (KZ) problem \cite{KibbleCosm,Zurek}. One of the most important predictions of this theory is the Kibble-Zurek mechanism (KZM) [see \cite{Kibble} for a review] which explains the formation and the density of topological defects in the off-equilibrium regime across a phase transition slowly driven by temperature. The Kibble-Zurek approach well describes the off-equilibrium dynamics near the transition and leads to a non-trivial scaling theory of the observables in terms of  appropriate length and time scales, different from those at the equilibrium. The scaling relations depend on the equilibrium critical exponents and also on some general features of the time-dependence of the external fields. In the limit of quasi-adiabatic time-variations $t_s\rightarrow\infty$, the results are universal. Several experiments have investigated these off-equilibrium phenomena,  in particular checking the predictions for the abundance of topological defects arising from the off-equilibrium conditions across the critical temperature, as predicted by the KZM. The first experiments meanly involved superfluids and superconductors. Modern proves of these behaviors principally come from cold-atoms experiments, ion crystals and from improved experiments still based on superfluids or superconductors [e.g. \cite{Lamporesi, TopCoulomb,Supercurr}]. In particular, Bose–Einstein condensates in trapped cold gases are extremely controllable systems and therefore an ideal platform to check the KZ mechanisms.\\
Off-equilibrium behaviors characterize also the first-order phase transitions (FOTs). In particular, one of the early off-equilibrium phenomena observed was the hysteresis, which has been widely studied [e.g. \cite{Binder,Rao,DharThomas,DharThomas2}]. Hysteresis arises in ferromagnetic systems when there is an external magnetic field with a time-dependence, such as $h(t,t_s)\approx t/t_s$, at fixed temperature. The magnetic field changes direction crossing the transition at $h=0$ followed by the magnetization  that also has to change direction according with $h$. However, the system reacts in late to the external perturbation developing metastable states for a certain interval of time. If we vary the external magnetic field cyclically across FOTs, these memory effects lead the system to dissipate a non-zero value of energy. The magnetic work done by the system can be valuated as $\oint dh(t,t_s) \cdot \Sigma(t,t_s)\propto \oint dt \cdot \Sigma(t,t_s)$ where $\Sigma$ is the magnetization of the system. The last integral is called hysteresis loop area and can be used to esthimate how far a system is from the equilibrium. Infact, if the magnetization presents the equilibrium behavior, the hysteresis curve is shrinked to a single line and the system does not spend energy in the cycle.\\
\\
The paper is organized as follows. In the Sec.\ref{SEC2} we introduce the $N$-vector model and its continuum limit as an euclidean field theory. The Sec.\ref{SEC3} is dedicated to the dynamics of the fields: we assume a relaxational dynamics for the model implemented by a purely dissipative Langevin equation with a white gaussian noise. In Sec.\ref{SEC4} we consider a time dependence also for the external magnetic field coupled to the system defining the examined protocol opportunely. The Sections \ref{SEC5} and \ref{SEC6} discuss the passage across the continuous phase transition while in Sec.\ref{FOT} the case $T<T_c$ is analyzed pointing out the analogies with the case $T=T_c$. In particular, the Sec.\ref{SEC5} provides to an introduction of the off-equilibrium scenario. The off-equilibrium scaling limit is defined, the finite-size effects are briefly discussed and the thermodynamic infinite-volume limit justified.
The large $N$ limit of the model is discussed in the Sec.\ref{SEC6}. The limit $N\rightarrow\infty$ allows us to derive analytical results. The scaling relations and the scaling functions for the correlators in the off-equilibrium regime across the continuous phase transition are investigated in the Sec.\ref{SEC7}.
 In Sec.\ref{FOT} we report the general scaling theory appropriate to describe the off-equilibrium and we consider the effects of a relaxational dynamics below the critical temperature. In the Sec.\ref{SEC9} a magnetic field protocol at $T<T_c$ is considered and the off-equilibrium behaviors for the O$(N)$ vector models are investigated. For both the cases $T\leq T_c$, in Sec.\ref{SEC10} we study the hysteresis phenomenon and derive the scaling relation of the magnetic work. 
In Sec.\ref{SEC11} we also discuss the asymptotic behaviors for the cases $T\leq T_c$ i.e. what happens when the system approaches the off-equilibrium regime. If the system reaches the equilibrium asymptotically, it is expected that the first deviations from the equilibrium background are exponentially damped. Finally, in Sec.\ref{SEC12} we draw some conclusion.\\

 \section{O$(N)$ vector model}\label{SEC2}
 
 The $N$-vector model is a lattice model where on each site $i$ lies an $N$-vector spin variable $S_i$ of unit length interacting through a short range ferromagnetic O$(N)$ symmetric two-body interaction $V_{i,j}$. The partition function of such a model can be written as \cite{ZJbook}
\begin{equation}\label{micromodel}
Z=\int \prod_i dS_i\cdot \delta(S^2_i -1)\cdot e^{-\varepsilon(S)/T},
\end{equation} 
in which the configuration energy $\varepsilon$ is
\begin{equation}\label{microE}
\varepsilon (S) = -\sum_{i,j} V_{i,j} ( S_i \cdot S_j)
\end{equation}
This model has a second order phase transition between a disordered phase at high temperature, and a low temperature ordered phase where the O$(N)$ symmetry is spontaneously broken, and the order parameter $S_i$ has a non-vanishing expectation value. One can add to $\varepsilon(S)$ a linear coupling
\begin{equation}\label{N-vec}
\varepsilon(S)=- \sum_{i,j} V_{i,j} ( S_i \cdot S_j)+\sum_i  h\cdot S_i
\end{equation}
which can be interpreted as a uniform external magnetic field. The presence of a non-zero magnetic field leads to a first-order transition in the low-temperature phase along the line of $h=0$. At the continuous transition the correlation length diverges and therefore a non-trivial long distance physics emerges.\\
The long distance physics of $\eqref{N-vec}$ can be described through an euclidean field theory, [see e.g.  \cite{ZJbook,LargeNzinn}], with action
\begin{equation}\label{action}
 S[\phi]=\int [ \frac{1}{2}(\partial_{\mu}\phi(x))^2+ \frac{1}{2}r\phi^2(x) +\frac{u}{4!}(\phi^2(x))^2-h_{\alpha}\phi_{\alpha}(x)]\cdot d^dx .
\end{equation}
where the spatial dimensions $2<d<4$.

 \section{Dynamics of the fields}\label{SEC3}
 We confer to the system a relaxational type of dynamics in which the fields satisfy a purely dissipative Langevin equation of motion:
 \begin{equation}\label{Lang}
 \partial_t \phi_{\alpha}(x,t) = -\frac{\Omega}{2} \frac{\delta S[\phi]}{\delta\phi_{\alpha}(x,t)} +\varsigma_{\alpha}(x,t)
 \end{equation}
 where we have inserted a white gaussian type of noise
 \begin{equation}
 <\varsigma_{\alpha}(x,t)>_{\varsigma}=0
 \end{equation}
 \begin{equation}
 <\varsigma_{\alpha}(x,t)\cdot \varsigma_{\beta}(x',t')>_{\varsigma}= \Omega\cdot \delta_{\alpha\beta}\cdot \delta(t-t')\delta^d(x-x')
 \end{equation}
This type of dynamics reproduces the effects of an heat-bath with which the system is in contact and leads the system to the equilibrium regime in the limits $t\rightarrow\pm\infty$. In the following we set $\Omega=2$ in order to recover the equilibrium propagator with the standard normalization after long times.

 \section{Protocols: dynamics of the parameters}\label{SEC4}
 We consider a time-dependence for the external fields coupled to the model. In particular we analize the case of an external magnetic field:
 \begin{equation}\label{protocol}
 h_{\alpha}(t,t_s)=\delta_{1\alpha}\cdot h(t,t_s) = \delta_{1\alpha} t/t_s
 \end{equation}
at the critical temperature $T=T_c$ and below it $T<T_c$. The time-scale of the variations of such magnetic field is  $t_s$ and we investigate the limit of very slow passage across the critical point $t_s\rightarrow\infty$. Thermal protocol for the underlying model $r(t,t_s)-r_c \approx -t/t_s$ at $h=0$ has been already widely discussed, e.g. \cite{CEGS-12}.\\
\\
Let us begin with the continuous phase transition occurring at $T=T_c$ and then we extend the formalism also to the first-order phase transition at $T<T_c$.\\
\newpage
\hrule
\begin{center}
\textbf{Case}  $T=T_c $
\end{center}
\hrule
\section{Off-equilibrium scaling regime}\label{SEC5}
 We assume that the system admits a non-trivial rescaling for $h(t,t_s)\simeq 0$ by measuring the time and the length scales through the Kibble-Zurek (KZ) scales [see \cite{Kibble,CEGS-12} for a review]:
 \begin{equation}\label{KZ}
 t_Q= (t_s)^{\nu_g z/(\nu_g z+1)} \qquad  l_Q= t_Q^{1/z},
 \end{equation}
 where $z$ is the dynamical critical exponent associated to the dynamics of the fields $\eqref{Lang}$ and $\nu_g$ is a generalized exponent \cite{CEGS-12} driving the divergence of the instantaneous equilibrium correlation length $\xi(t,t_s)$ in terms of the protocol close to the critical point $h(t,t_s)\simeq 0$. For a magnetic field protocol, $\nu_g$ is given by:
 \begin{equation}
 \nu_g= 1/d_h= \frac{2}{d+2-\eta} \end{equation}
 Thus,
 \begin{equation}
 \xi(t,t_s) \sim |t/t_s|^{-\nu_g}\end{equation}
 We can define also the instantaneous equilibrium relaxation time of the system $\xi_t(t,t_s)$ which diverges close to the critical point as
 \begin{equation}
 \xi_t(t,t_s) \sim |t/t_s|^{-z\nu_g}\end{equation}
 because of the critical slowing down phenomenon.\\
The off-equilibrium scaling limit is the limit $t_s\rightarrow \infty$ holding $t/t_Q$ and $x/l_Q$ fixed where the correlation functions of the system exhibit a behavior:
 \begin{equation}\label{scalingoff}
 G^{(n)}(x_1,\cdots,x_n,t) \sim l_Q^{-\alpha}\cdot \mathcal{G}^{(n)}(x_1/l_Q, \cdots, x_n/l_Q, t/t_Q)
 \end{equation}
 \\
 where $\alpha$ is the scaling dimension and $\mathcal{G}^{(n)}$ is the off-equilibrium scaling function associated to the $n$-points correlation function.\\
The off-equilibrium physics is observed for an interval of size $t_Q\sim (t_s)^e$  around the transition, with $e=(z/d_h)/(1+z/d_h) < 1$. When we take the limit $t_s\rightarrow \infty$, this interval becomes very large $t_Q\rightarrow \infty$. In terms of the KZ scales the protocol is given by
\begin{equation}
h(t,t_s)\sim t/t_Q \cdot l_Q^{-d_h} \overset{t_s\rightarrow\infty}{\rightarrow} 0.
\end{equation}
Since $e<1$ the focus on the off-equilibrium scenario is related to a very small values of the external field. The dynamics presents universal scaling behaviors in the limit $t_s\rightarrow \infty$ depending only on the equilibrium static and dynamical critical exponents plus the exponent $\nu_g$. It does not depend on the choice of the initial and final value of the magnetic field because the off-equilibrium scaling occurs in a range of values of the protocol that shrinks near zero when $t_s\rightarrow \infty$.

 \subsection*{Finite-size effects and infinite-volume limit}
 Let us discuss the off-equilibrium scaling arising by the presence of time-dependent magnetic field $\eqref{protocol}$ coupled to a system of finite size $L$ which approaches the critical point. Assuming the exsistence of a non-trivial scaling for $h(t,t_s)\simeq 0$, we expect that the off-equilibrium behavior is controlled by the scaling variables:
 \begin{equation}
\overline{t}=t/t_Q, \qquad \overline{x}=x/l_Q, \qquad \ell=l_Q/L.
 \end{equation}
The magnetic field can be written as
\begin{equation}
h(t,t_s) = t/t_s=\overline{t}\cdot \ell^{-1/\nu_g}.
\end{equation}
Since the system is at the instantaneous equilibrium for a configuration of the external field $h(t,t_s)$, a statistical observable $O$ with scaling dimension $\Delta$ presents a finite-size scaling behavior:
\begin{equation}\label{finitesizeq}\begin{split}
<O(x,t,h,L)O(0,t,h,L)>_{\varsigma}=G_{OO}(x,t,h,L) \sim\\
 L^{-2\Delta}\cdot \mathcal{G}_{OO}^{\text{eq.}}(\frac{x}{L}, \frac{t}{L^z},h\cdot L^{1/\nu_g}),
 \end{split}
\end{equation}
when $|h|\rightarrow 0$, $ \mathcal{G}_{OO}^{\text{eq.}}$ is the two-point equilibrium correlator scaling function. The brackets above mean an expectation value performed over the noise distribution. In the infinite-volume limit $L\rightarrow \infty$, the equation $\eqref{finitesizeq}$ becomes:
\begin{equation}\begin{split}
<O(x,t,h,L)O(0,t,h,L)>_{\varsigma} \sim \xi^{-2\Delta}\cdot \mathcal{G}_{OO}^{\text{eq.}}(\frac{x}{\xi}, \frac{t}{\xi^z})\\
 = |h|^{2\nu_g\Delta}\cdot  \mathcal{G}_{OO}^{\text{eq.}}(x|h|^{\nu_g},t|h|^{z\nu_g}),
\end{split}\end{equation}
for $|h|\rightarrow 0$ at $x/\xi$, $t/\xi^z$ fixed.\\
We require that the equilibrium finite size scaling matches the infinite-volume behavior 
\begin{equation}\begin{split}
\mathcal{G}_{OO}^{\text{eq.}}(\frac{x}{L}, \frac{t}{L^z},h\cdot L^{1/\nu_g}) \overset{L=\infty}{\sim}\\
 |h\cdot L^{1/\nu_g}|^{2\nu_g\Delta}\cdot \mathcal{G}_{OO}^{\text{eq.}}(x|h|^{\nu_g},t|h|^{z\nu_g}).
 \end{split}\end{equation}
 when $|h|\cdot L^{1/\nu_g}\rightarrow \infty$.\\
 Let us now define the finite-size off-equilibrium scaling limit as the limit $t_s$, $L\rightarrow\infty$ keeping the variables $\overline{t}$, $\overline{x}$ and $\ell$ fixed. In this limit, the correlator $G_{OO}$ has a rescaling:
 \begin{equation}
 G_{OO}(x,t,h,L) \sim L^{-2\Delta}\cdot \mathcal{G}_{OO}(\overline{x},\overline{t},\ell).
 \end{equation}
where $\mathcal{G}_{OO}$ is a general function of the off-equilibrium scaling variables. The infinite-volume limit can be obtained by performing the limit $\ell\rightarrow 0$ at fixed $\overline{t}$, $\overline{x}$:
\begin{equation}
\mathcal{G}_{OO}(\overline{x},\overline{t},\ell) \overset{\ell\rightarrow 0}{\sim} \ell^{-2\Delta} \cdot \mathcal{G}_{OO}(\overline{x},\overline{t}).
\end{equation}
The previous relation states that the Eq.$\eqref{scalingoff}$ is well-defined. Note that the results above apply for general protocols and are not specific for an external magnetic field at $T=T_c$. In the following, we consider the infinite-volume limit. Similiar arguments for the finite-size off-equilibrium scaling and its infinite-volume limit are reported in \cite{VicOff}.

 \section{Large $N$ limit}\label{SEC6}
 We consider the O$(N)$ vector model in the limit of large $N$, where it allows analytical computations.\\
 It is well known that the system in the limit of large $N$ has zero anomalous dimensions of the fields [e.g. \cite{ZJbook, LargeNzinn, LargeN2}]:
 \begin{equation}
 \eta=0 \Rightarrow \begin{cases}  d_{\phi}=(d-2)/2 \\
 \nu_g=1/d_h= 2/(d+2)\\
 z=2\\
 \end{cases}
 \end{equation}
  where $d_{\phi}=(d-2+\eta)/2$ is the scaling dimension of the fields. Furthermore, the system obeys to a set of  equations at the leading order in $1/N$ called saddle point equations. The crucial point of the simplification introduced by the large $N$ limit relies upon the definition of an effective mass term $m^2$:
 \begin{equation}
 m^2-r-\frac{u}{6} <\phi^2>_{\varsigma} =0;
 \end{equation}
 At the leading order in $1/N$, one can neglect the variations of the square fields $\phi^2(x,t)\sim <\phi^2(x,t)>_{\varsigma}$ and consider the theory as quasi-gaussian with a mass $m^2$. 

 \subsection*{Equation of motion at large $N$}
In the large $N$ limit it is possible to linearize the Langevin equation by introducing the time-dependent effective mass term $m^2(t,t_s)$:
\begin{equation}\begin{split}
\partial_t \phi_{\alpha}(x,t) \overset{N\rightarrow \infty}{\simeq} -\Big(-\partial_{\mu}\partial_{\mu} +m^2(t,t_s) \Big) \phi_{\alpha}(x,t) 
 \\
 + h(t,t_s) +\varsigma_{\alpha}(x,t);
\end{split}\end{equation}
The solution of the equation above, written in the Fourier transform is given by:
 \begin{equation}\begin{split}
 \phi_{\alpha}(k,t) = \phi^0_{\alpha}(k,t)+ \int_{t_0}^t dt'\cdot \exp\Big(-\int_{t'}^t dt''\cdot (k^2+\\
 m^2(t'',t_s))\Big) \cdot \{ (2\pi)^d \delta^d(k) \delta_{1\alpha} h(t',t_s)+ \varsigma_{\alpha}(k,t')\}
 \end{split} \end{equation}
  where
 \begin{equation}\label{phi0}\begin{split}
 \phi^0_{\alpha}(k,t)=\exp\Big(-\int_{t_0}^t dt'\cdot (k^2+m^2(t',t_s))\Big) (2\pi)^d \delta^d(k) \delta_{1\alpha} \sigma
 \end{split}\end{equation}
 and $\sigma$ is the equilibrium expectation value of the field.
 
 \subsection*{Correlation functions}
 
From the solution for the field, it is possible to extract the value of the one and two-point correlation functions. Let us begin with the magnetization of the system $\Sigma(t,t_s)$:
\[
 <\phi_{\alpha}(k,t)>_{\varsigma}= (2\pi)^d \delta^d(k)\delta_{1\alpha} \Sigma(t,t_s)\]
 \begin{equation}\label{magn}\begin{split}
 \Sigma(t,t_s)=\Sigma^0(t,t_s) + \int_{t_0}^t dt'\cdot  h(t',t_s) \cdot \\
 \exp\Big(-\int_{t'}^t dt''\cdot m^2(t'',t_s) \Big) \end{split}
\end{equation}
 with
 \begin{equation}
 \Sigma^0(t,t_s) = \sigma\cdot \exp\Big(-\int_{t_0}^t dt'\cdot m^2(t',t_s)\Big)
 \end{equation}
 The expectation value of two fields is given by:
 \begin{equation}
 <\phi_{\alpha}(k,t)\phi_{\beta}(k',t)>_{\varsigma}= \delta_{\alpha\beta} (2\pi)^d \delta^d(k+k') G_T(k,t,t_s)
 \end{equation}
 for the transverse components $\alpha>1$, $\beta>1$. It follows that the transverse two-point correlation function is:
 \begin{equation}\label{G_T}
 G_T(k,t,t_s)=2\int_{t_0}^{t}dt'\cdot \exp\Big(-2\int_{t'}^t dt''\cdot (k^2 +m^2(t'',t_s))\Big)
 \end{equation}
 The longitudinal two-point correlation function is given by:
 \[
 <\phi_{1}(k,t)\phi_{1}(k',t)>_{\varsigma}=(2\pi)^d \delta^d(k+k') G_L(k,t,t_s)\]
 \[
 G_L(k,t,t_s) = G^0(k,t,t_s) + (2\pi)^d \delta^d(k) \Sigma^2(t,t_s)+\] \begin{equation}\label{G_L}
 2\int_{t_0}^{t}dt'\cdot \exp\Big(-2\int_{t'}^t dt''\cdot (k^2 + m^2(t'',t_s))\Big)\end{equation}
 with 
 \[
 G^0(k,t,t_s)= (2\pi)^d\delta^d(k)\Big[ (\Sigma^0(t,t_s))^2+\]
 \begin{equation}
 2\Sigma^0(t,t_s)\int_{t_0}^t dt'\cdot \exp\Big(-\int_{t_0}^t dt''\cdot (k^2+m^2(t'',t_s)) \Big) h(t',t_s) \Big]
 \end{equation}
  
  \subsection*{Constraint equations}
 The large $N$ limit gives a set of dynamical equations \cite{MazZann} which relate the quantities of interest to the effective mass term of the model $m^2(t,t_s)$:
 \begin{equation}\label{a}
 \partial_t \Sigma(t,t_s) = -m^2(t,t_s)\cdot \Sigma(t,t_s) + h(t,t_s);\end{equation}
  \begin{equation}\label{b}
 \partial_t G_T(k,t,t_s)= -2(k^2+m^2(t,t_s))\cdot G_T +2;\end{equation}
 and finally the consistence equation which defines the effective mass term at any instant of time:
 \begin{equation}\label{c}
 r+\frac{u}{6}\Big(\Sigma^2(t,t_s) +\int^{\Lambda} \frac{d^dk}{(2\pi)^d} \cdot G_T(k,t,t_s)\Big)=m^2(t,t_s).\end{equation}
 where $\Lambda$ is an UV cutoff. Note that the last equation complete the theory: the dynamical correlators are function of the external magnetic field (whose dynamics is given at all times by the protocol) and of the effective mass term. Solving this equation for $m^2(t,t_s)$ we can know the behavior of the correlators at any instant of time.
 
\section{Scaling relations}\label{SEC7}
There is an important thing to point out: the role of the initial condition in the KZ scaling limit. Since the scaling behaviors turn out to be universal in the limit of slow variations $t_s\rightarrow\infty$, it is expected that the initial state of the system at $t_0$ does not influence the critical theory. The correlators $\eqref{magn}$, $\eqref{G_T}$ and $\eqref{G_L}$, depend on $\phi^0$ through $\Sigma^0$ and $G^0$. The focus on the off-equilibrium scenario enlarge the small area near $|t|\sim 0$ making the starting time $t_0$ very far. Therefore, the term $\eqref{phi0}$ is exponentially driven to zero and the sensibility of the system on the initial state dissappears. 
 
 Let us compute the scaling relations for the correlators in the off-equilibrium regime. We start from the constraint Eq.$\eqref{c}$ in the case $T=T_c$ i.e. $r=r_c$ with
\begin{equation}
r_c= -\frac{u}{6}\int^{\Lambda} \frac{d^dk}{(2\pi)^d} \cdot G_T(k,t,t_s,m^2=0)
\end{equation}
 Thus, the Eq.$\eqref{c}$ can be written as
  \begin{equation}\begin{split}
 m^2(t,t_s)= \frac{u}{6}\Big( \Sigma^2(t,t_s) + \int^{\Lambda} \frac{d^dk}{(2\pi)^d}\cdot \\
 \Big[ G_T(k,t,t_s,m^2)-G_T(k,t,t_s,0)\Big]\Big)
 \end{split}
 \end{equation}
 
 We consider the KZ scaling limit $t_s\rightarrow \infty$ keeping $\overline{t}$ and $k\cdot l_Q =\overline{k}$ fixed. We make the following scaling hypotesis:\\
\\
$\circ$ $m^2(t;t_s) \sim \mathcal{M}^2(\overline{t}) \cdot l_Q^{-2}$, using dimensional arguments \cite{CEGS-12}.\\
\\
$\circ$ $G_{T,L}(k,t,t_s) \sim \mathcal{G}_{T,L}(\overline{k},\overline{t}) \cdot l_Q^2$.\\
\\
$\circ$ $\Sigma(t,t_s) \sim \Theta(\overline{t}) \cdot l_Q^{-d_{\phi}}$.\\
\\
$\circ$ $h(t,t_s)= t/t_s \sim \overline{t} \cdot l_Q^{-d_h}$\\
\\
where $\Theta$, $\mathcal{G}_{T,L}$ are the off-equilibrium scaling function of the magnetization and of the two-point correlators respectively. One can verify the scaling hypotesis for the correlation functions starting from the Eq.$\eqref{magn}$ and $\eqref{G_T}$, $\eqref{G_L}$ finding:

\begin{equation}\label{magnscal}
\Theta(\overline{t})=\int_{-\infty}^{\overline{t}} d\overline{t}'\cdot \overline{t}'\cdot \exp\Big(-\int_{\overline{t}'}^{\overline{t}} d\overline{t}'' \cdot \mathcal{M}^2(\overline{t}'') \Big)
\end{equation}
 for the scaling function of the magnetization and
 \begin{equation}\label{GTscal}
 \mathcal{G}_T(\overline{k},\overline{t})= 2\int_{-\infty}^{\overline{t}} d\overline{t}'\cdot \exp\Big(-2\int_{\overline{t}'}^{\overline{t}} d\overline{t}''\cdot (\overline{k}^2 +\mathcal{M}^2(\overline{t}))\Big)
 \end{equation}
 for the scaling function of the transverse two-point correlator. The scaling function of $G_L$ is given by 
\begin{equation}\label{GLscal}\begin{split}
 \mathcal{G}_L(\overline{k},\overline{t}) =(2\pi)^d \delta^d(\overline{k}) \cdot \Theta^2(\overline{t})+\\ 2\int_{-\infty}^{\overline{t}} d\overline{t}'\cdot \exp\Big(-2\int_{\overline{t}'}^{\overline{t}} d\overline{t}''\cdot (\overline{k}^2 +\mathcal{M}^2(\overline{t}))\Big)
\end{split}\end{equation}
 At this point, we can consider the KZ scaling limit in the Eq.$\eqref{c}$ [see also \cite{CEGS-12}]
 \begin{equation}\label{cscal}\begin{split}
 \mathcal{M}^2(\overline{t})\cdot l_Q^{-2} \sim 0 =l_Q^{-(d-2)} \cdot \frac{u}{6}\Big(\Theta^2(\overline{t}) +\\
 \int^{+\infty} \frac{d^d\overline{k}}{(2\pi)^d} \cdot \Big[\mathcal{G}_T(\overline{k},\overline{t}, \mathcal{M}^2) -\mathcal{G}_T(\overline{k},\overline{t},0)\Big] \Big)
 \end{split}\end{equation}
 where we have set $m^2$ to zero because is a subleading term in the KZ scaling limit.\\
We have explicitly computed the KZ scaling functions of the correlators $\eqref{magnscal}$, $\eqref{GTscal}$ and $\eqref{GLscal}$: these quantities depend on the scaling function of the effective mass term $\mathcal{M}^2$ which is defined at any instant of time as the solution of the Eq.$\eqref{cscal}$. Unfortunately, the Eq.$\eqref{cscal}$ does not have known solution.\\
Let us now investigate the limit $t_s\rightarrow\infty$ holding $\overline{t}=t/(t_s)^e$ and $\overline{k}=k\cdot (t_s)^{e/z}$ fixed, with $e$, $z$ free parameters. We show that there is only a value of $e$ and $z$ which lead to a non-trivial rescaling. The transverse two-point correlation function has a rescaling:
\begin{equation}\label{rescalGT}
G_T(k,t,t_s) \sim (t_s)^e \cdot \mathcal{G}_T(\overline{k},\overline{t})
\end{equation}
where we have rescaled $m^2(t,t_s)\sim \mathcal{M}^2(\overline{t})\cdot (t_s)^{-2e/z}$ because of a dimensional analysis. There is only a choice of $z$ which make the exponential in Eq.$\eqref{G_T}$ a scaling quantity:
\begin{equation}
-2e/z + e=0 \Rightarrow z=2.
\end{equation}
The rescaling of the magnetization is:
\begin{equation}\label{rescalM}
\Sigma(t,t_s) \sim (t_s)^{2e-1} \cdot \Theta(\overline{t}).
\end{equation}
Performing the off-equilibrium scaling limit in the Eq.$\eqref{cscal}$, we obtain:
\[
 (t_s)^{-e} \cdot \mathcal{M}^2(\overline{t}) = \frac{u}{6}\Big((t_s)^{4e-2}\cdot \Theta^2(\overline{t}) +\]
 \begin{equation}
 \int^{+\infty} \frac{d^d\overline{k}}{(2\pi)^d} \cdot (t_s)^{-de/z}  \cdot (t_s)^e \cdot \Big[\mathcal{G}_T(\overline{k},\overline{t}, \mathcal{M}^2) -\mathcal{G}_T(\overline{k},\overline{t},0)\Big] \Big)
 \end{equation}
 The mass term is subleading. The equation above leads to a non-trivial rescaling only if
 \begin{equation}
 -\frac{d}{z}e+e=4e-2 \Rightarrow e=\frac{2z}{(d+3z)} \overset{\text{?}}{=} \frac{\nu_g z}{(\nu_g z +1)}
 \end{equation}
 These results are in agreement with the definition of the KZ scales $\eqref{KZ}$ setting $\nu_g=1/d_h$ in $d=3$.\\
 We conclude that the KZ scaling is the only one that the system admits in the off-equilibrium regime.
\newpage
\hrule
\begin{center}
\textbf{Case}  $T<T_c $
\end{center}
\hrule

 \section{First-order phase transition}\label{FOT}
 The general features of the off-equilibrium behaviors are not specific of the continuous phase transition: the same scaling theory, with appropriate exponents, can be applied also below the critical temperature, where the O$(N)$ vector model undergoes a first-order phase transition (FOT) along the line of $h=0$.\\
 We investigate  therefore the FOT driven by the magnetic field $\eqref{protocol}$ around $h=0$ at fixed temperature $T<T_c$. \\
 The first step consists in the construction of a scaling theory also for the FOTs. By following the works \cite{scalfirst,PF-83,FP-85} we have the results:
 \begin{equation}
 d_{\phi}=0 \qquad d_h=d 
 \end{equation}
 for the scaling dimensions of the fields and of the magnetic field respectively. \\
 Across a FOT the system switches from one non-critical phese, with finite correlation length, to another non-critical phase. However, when $h\rightarrow 0$ a long-range order arises because the system does not distinguish the ordered phases anymore \cite{ScalNew}. Thus, even if the correlation length is still finite, we associate to the FOT a coherence length $\xi(t,t_s)$ which diverges for $h(t,t_s)\simeq 0$ as
\begin{equation}
\xi(t,t_s) \sim |h(t,t_s)|^{-1/d_h} = |t/t_s|^{-1/d}
\end{equation}
and it follows that
\begin{equation}
\nu_g= 1/d_h=1/d
\end{equation}
 for a magnetic field protocol also across FOTs.
 \subsection*{Relaxational dynamics at $T<T_c$}
 We assume that the dynamics of the fields occurs through a purely dissipative Langevin equation $\eqref{Lang}$ also below the critical temperature. It is useful to consider a system of finite-size $L$ with a cubic shape $V=L^d$ and focus on the case $N=2$. Infact, as will be more clear in the following, the results apply for each $N\geq 2$.\\
 We report the results for the dynamical exponents of the ref.\cite{VicOff}. Firstly, let us consider a spin system without magnetic fields:  in this case the magnetization has fixed modulus but there are no constraints on the direction. The random orientation of the vector magnetization into the space is expected to be the slowest dynamics of the system and has a dynamical exponent $z=d$ because requires a variation in the enteire volume of the system. There is also a motion in the transverse planes due to the spin-waves with a time-scale $\sim L^2$ ($z=2$). However, it is faster and thermalizes over larger time-scales $\sim L^d$.\\
In the presence of a magnetic field, the magnetization has a fixed direction. The only degrees of freedom of the system are the spin-waves propagating along the transverse directions.\\
Therefore we have the follwing scenario for the dynamics exponents \cite{VicOff}:
\begin{equation}
T< T_c\begin{cases}
z=d \qquad \text{if}\qquad h=0\\
z=2 \qquad \text{if} \qquad h\neq 0
\end{cases}
\end{equation}
 \section{Off-equilibrium scaling across the FOT}\label{SEC9}
Let us extract the off-equilibrium scaling relations arising across the FOT in the O$(N)$ vector model at large $N$. The constraint equation $\eqref{c}$ is still valid below the critical temperature because related to the large $N$ limit and not to the specific protocol. Furthermore, since we assume a relaxational dynamics of the fields, we can use the results $\eqref{magn}$, $\eqref{G_T}$ and $\eqref{G_L}$ for the correlation functions in the presence of a time-dependent magnetic field $\eqref{protocol}$. These are function of the effective mass term which is defined at all times by the equation:

\begin{equation}\label{clow}
r+\frac{u}{6}\Big(\Sigma^2(t,t_s) +\int^{\Lambda}\frac{d^dk}{(2\pi)^d} \cdot G_T(k,t,t_s) \Big) = m^2(t,t_s).
\end{equation}
One can note that the specific value of temperature $T<T_c$ does not influence the off-equilibrium scaling because the results of Sec.\ref{FOT}  apply for each value of the temperature $T<T_c$ (but are quite different from the case $T=T_c$).\\ 
We consider the off-equilibrium scaling limit in the constraint equation above: we take the limit $t_s\rightarrow\infty$ keeping $\overline{t}=t/(t_s)^e$ and $\overline{k}=k\cdot (t_s)^{e/z}$ fixed. Since we vary the magnetic field at fixed temperature, the thermal coupling $r$ is a constant and scales as $r\sim r\cdot (t_s)^0$. Thus, from the rescaling of the magnetization $\eqref{rescalM}$, we have
\begin{equation}
2(2e -1) =0 \Rightarrow e= \frac{1}{2}, \qquad \forall d.
\end{equation}
The exponent $e$ in the low-temperature phase does not depend on the spatial dimension of the system. We define the off-equilibrium time-scale $t_Q$ across the FOT as
\begin{equation}\label{FOTtQ}
t_Q= (t_s)^e= \sqrt{t_s}
\end{equation}
The off-equilibrium scaling limit of the constraint Eq.$\eqref{clow}$ requires that the magnetization scales as $(t_s)^0$ i.e. that the scaling dimensions $d_{\phi}=0$, in agreement with the relations of Sec.\ref{FOT}. 
The off-equilibrium dynamics arises for very small values of the magnetic field,
\begin{equation}
h(t,t_s)= t/t_s= (t/\sqrt{t_s})\cdot (t_s)^{-1/2} \overset{t_s\rightarrow \infty}{\rightarrow} 0.
\end{equation}
Therefore, the slowest dynamics of the system in the off-equilibrium regime is expected to be a change in the direction of the vector magnetization whose time-scale is given by the dynamical exponent $z=d$. Other types of dynamics occur with faster time-scales and can be neglected. It follows that the off-equilibrium length scale $l_Q$ across the FOT can be defined as
\begin{equation}\label{FOTlQ}
l_Q=t_Q^{1/z}=t_Q^{1/d}=(t_s)^{1/2d}.
\end{equation}
For the rescaling of the transverse two-point correlation function, we read into the Eq.$\eqref{rescalGT}$ that
\begin{equation}
-(d-z)/z=0 \Rightarrow z=d
\end{equation}
The two-point correlations among the transverse components are leading terms only if we consider the slowest dynamics of the system.\\
The name KZ generally refers to the case of the off-equilibrium dynamics across continuous phase transitions. However, we can note that the definition of the off-equilibrium scales $\eqref{FOTtQ}$, $\eqref{FOTlQ}$ are in agreement with $\eqref{KZ}$ setting $\nu_g=1/d$ and $z=d$.\\
The effective mass term satisfies the constraint Eq.$\eqref{a}$ also below the critical temperature: in the off-equilibrium scaling limit, Eq.$\eqref{a}$ becomes
\begin{equation}\begin{split}
m^2(t,t_s)\sim -\frac{\frac{d}{d\overline{t}}\Theta(\overline{t})}{\Theta(\overline{t})} \cdot t_Q^{-1}+ \frac{\overline{t}}{\Theta(\overline{t})}\cdot l_Q^{-d_h}\\ =-\frac{\frac{d}{d\overline{t}}\Theta(\overline{t})}{\Theta(\overline{t})}\cdot t_Q^{-1}+\frac{\overline{t}}{\Theta(\overline{t})}\cdot t_Q^{-d/z}\end{split}
\end{equation}
Therefore the scaling relation of the effective mass term is given by
\begin{equation}
m^2(t,t_s) \sim \mathcal{M}^2(\overline{t}) \cdot l_Q^{-d}.
\end{equation}
Note that the transverse two-point correlation function generally has a critical behavior in the limit of zero momenta. In this case the transverse correlation function remains finite at all times because
\begin{equation}\label{scaling mass}
m^2(t,t_s) \sim 0 \cdot l_Q^{-2} + \mathcal{M}^2(\overline{t})\cdot l_Q^{-d}.
\end{equation}
The presence of a small magnetic field makes the transverse susceptibility not IR-divergent.\\ 
Let us discuss the rescaling of the momenta. Away from the transition, we can roughly approximate the magnetization as a constant $\Sigma(t,t_s)\approx \sigma$ and using the Eq.$\eqref{a}$:
\begin{equation}\label{meq}
m^2(t,t_s) \approx \frac{t}{t_s \sigma}.
\end{equation}
We know that this approximation breaks down when $t\simeq 0$. However, the Eq.$\eqref{meq}$ permits to compute explicitly the correlation function. In particular, we study the transverse two-point correlation function $\eqref{G_T}$ \cite{DharThomas}:
\begin{equation}
G_T (k,t,t_s)  \approx e^{s^2}\cdot \sqrt{\pi|\sigma|  t_s}\cdot \text{Erfc}(s)
\end{equation}
where we have defined the variable $s=(k^2 t_s|\sigma| + |t|)/\sqrt{t_s|\sigma|}$. Keeping the leading part for $s$ large, we obtain:
\begin{equation}
 G_T(k,t,t_s) \sim \frac{\sqrt{|\sigma| t_s}}{s} \approx \frac{1}{k^2 +m^2(t,t_s)}.
\end{equation}
We recover the equilibrium value of the transverse two-point correlation function. Note that the limit $s\rightarrow \infty$ does not necessary imply $|\overline{t}|\rightarrow\infty$. Since we consider large-momenta, the system appear at instant thermal equilibrium at all the times [see also \cite{DharThomas, DharThomas2}]. In other words, below the critical temperature the degrees of freedom of the system are essentially given by the spin-waves. The off-equilibrium behavior depends only on these long-wavelength modes. There is a value of the momentum $k^{\diamond}$ which defines a cross-over behavior and separates the local-fluctuation regime $k>k^{\diamond}$ from the spin-waves $k<k^{\diamond}$. Its value can be computed self-consistently: if we assume that the large-momenta modes are at the instantaneous equilibrium for all times, the approximation $\eqref{meq}$  remains valid even when $t\rightarrow 0$. Thus, we can identify $k^{\diamond}$ as the value in which $s\sim O(1)$ at $t=0$ \cite{DharThomas}. It follows that
\begin{equation}
k^{\diamond} = (1/t_s |\sigma|)^{1/4}
\end{equation}
When we perform the off-equilibrium scaling limit, we enlarge the interval of the small momenta:  $k\cdot (t_s)^{1/2d}$ remains fixed when  $t_s\rightarrow \infty$. In particular, the rescaled value of $k^{\diamond}$ is
\begin{equation}
k^{\diamond}\cdot (t_s)^{1/2d} = (1/t_s |\sigma|)^{1/4} \cdot  (t_s)^{1/2d} \propto (t_s)^{-(d-2)/4d} \rightarrow 0
\end{equation}
because $2<d<4$. Since we consider very-low frequency protocols $t_s\rightarrow \infty$, the cross-over value of the momenta $k^{\diamond}$ tends to zero and this is equivalent to consider the momenta as subleading terms in the off-equilibrium scaling limit. It is useful to introduce the quantity:
\begin{equation}
S(t,t_s)=\int^{\Lambda} \frac{d^dk}{(2\pi)^d} \cdot G_T(k\rightarrow 0,t,t_s) \sim l_Q^0 \cdot \mathcal{S}(\overline{t});
\end{equation}
which summerizes the amount of the transverse correlations. Thus, in the off-equilibrium scaling limit, the constraint Eq.$\eqref{clow}$ can be written as
\begin{equation}\label{rescalclow}
r+\frac{u}{6}\Big(\Theta^2(\overline{t})+\mathcal{S}(\overline{t})\Big)=0;
\end{equation}
because the effective mass is subleading also below the critical temperature. The previous result $\eqref{rescalclow}$ can be written also as
\begin{equation}\label{clowrot}
\Theta^2(\overline{t}) +\mathcal{S}(\overline{t})=\sigma^2
\end{equation}
where $|\sigma|=\sqrt{-6r/u}$ is the equilibrium magnetization in the low-temperature phase. The Eq.$\eqref{clowrot}$ states that the vector magnetization performs a rigid rotation with fixed length equal to $|\sigma|$. The equilibrium behavior is recovered in the appropriate limits: away from the FOT, the magnetic field $h(t,t_s)\neq 0$ and the function $\mathcal{S}(\overline{t})$ is subleading. Therefore, the vector magnetization lies on the longitudinal direction given by the magnetic field and has a length $|\sigma|$. In contrast, in the off-equilibrium region, the longitudinal component of the magnetization first decreases dissipating into the transverse modes and then increases going to the opposite equilibrium value [see also \cite{Rao}]. Note that the transverse magnetization is zero because of the O$(N-1)$ symmetry but the correlations among the transverse components (resumed in the function $\mathcal{S}$) rotate the vector in one of the equally probable $N-1$ planes transverse to the longitudinal direction. Since the dynamics occurs into a plane, the system exhibits the same behavior for each $N\geq 2$ \cite{DharThomas2}.

\textbf{\hrule}
 \section{Hysteresis phenomena}\label{SEC10}
We define the hysteresis loop area $\mathcal{A}$ as the area beetween the two cuves described by the magnetization $\Sigma(t,t_s)$ going from  $t_i=-\infty$  to $t_f=+\infty$ and coming back (round-trip protocol $\gamma$) when the dynamics of the system is driven by the magnetic field protocol $\eqref{protocol}$  \cite{VicOff}:
\begin{equation}\label{area}
\mathcal{A}=\oint_{\gamma} dt\cdot \Sigma(t,t_s) 
\end{equation}
We note that the magnetization $\eqref{magn}$ has a symmetry with respect to a reflection of the magnetic field: if we reverse the direction of the magnetic field $h\mapsto -h$, it follows 
\begin{equation}
\Sigma^{\text{inv.}}(t,t_s) = -\Sigma(-t,t_s);
\end{equation}
The vaue of the magnetization with reversed time is:
\begin{equation}
\Sigma(-t,t_s) =\int_{-\infty}^{t} dt'\cdot \exp\Big(+\int_{t'}^{t} dt'' \cdot m^2(t'';t_s)\Big) \cdot t'/t_s.
\end{equation}
Thus, the hysteresis loop area can be also written as
\begin{equation}\label{hyst}\begin{split}
\mathcal{A}=\oint_{\gamma} dt\cdot \Sigma(t,t_s) =\int_{t_i=-\infty}^{t_f=+\infty} dt\cdot \Big( \Sigma(t,t_s)+\Sigma(-t,t_s) \Big)\\
=\int_{-\infty}^{+\infty} dt \cdot \int_{-\infty}^t dt' \cdot (t'/t_s) \cdot  2\cosh\Big(\int_{t'}^t m^2(t'',t_s)\cdot dt''\Big).
\end{split}
\end{equation}
The hysteresis loop area can be easly connected with the magnetic work $\mathcal{W}$ which the system performs over $\gamma$
\begin{equation}
\mathcal{W}=\oint_{\gamma} dh(t,t_s)\cdot \Sigma(t,t_s) = t_s^{-1} \cdot \mathcal{A}.
\end{equation}
Therefore the hysteresis loop area has a direct physical meaning. At the equilibrium the integral $\eqref{area}$ is equal to zero, so $\mathcal{A}$ quantifies how far the system is from the equilibrium.

\subsection*{Case $T=T_c$}
We consider the off-equilibrium scaling relation for the hysteresis loop area $\eqref{hyst}$ across the continuous phase transition. We perform the limit $t_s\rightarrow \infty$ at fixed $\overline{t}=t/t_Q$ and $\overline{k}=k\cdot l_Q$, where $t_Q$, $l_Q$ are given by $\eqref{KZ}$. The scaling relation of the hysteresis loop area is
\begin{equation}\label{hystscal}
 \mathcal{A} \sim l_Q^{2-d_{\phi}}\cdot \Xi.
\end{equation}
The amplitude $\Xi$ of hysteresis loop area is a constant which depends on the scaling function $\mathcal{M}^2$:
\begin{equation}\label{Xi}
\Xi=2 \int_{-\infty}^{+\infty}d\overline{t}\cdot\int_{-\infty}^{\overline{t}}d\overline{t}'\cdot \overline{t}' \cdot
\cosh\Big(\int_{\overline{t}'}^{\overline{t}} \mathcal{M}^2(\overline{t}'')\cdot d\overline{t}''\Big);
\end{equation}
From Eq.$\eqref{hystscal}$, it follows that the magnetic work has a scaling relation:
\begin{equation}
\mathcal{W}\sim (t_s)^{-2/3} \cdot \Xi
\end{equation}
in three spatial dimensions. This means that the energy spent by the system in a cycle decrease as $t_s$ becomes large. Hysteresis phenomena can be studied for an O$(3)$ vector model in $d=3$ : using the critical exponents of the Heisenberg universality class [see \cite{VicariRew}] in $\eqref{hystscal}$, we find a scaling relation
\begin{equation}
\mathcal{A} \sim (t_s)^{0.33}\cdot \Xi^{(N=3)},
\end{equation}
in agreement with the numerical results \cite{VicOff}. It is expected that the hysteresis loop area at large $N$ is qualitatively similiar to the case $N=3$. It follows that the magnetic work done by the system in a round-trip protocol for an O$(3)$ Heisenberg ferromagnet in $d=3$ scales as
\begin{equation}
\mathcal{W} \sim (t_s)^{-0.66} \cdot \Xi^{(N=3)}.
\end{equation}

\subsection*{Case $T<T_c$}
Let us compute the scaling relation for the hysteresis loop area also across the FOT. In particular, we take the limit $t_s\rightarrow\infty$ keeping $\overline{t}=t/t_Q=t/\sqrt{t_s}$ and $\overline{k}=k\cdot l_Q=k\cdot (t_s)^{1/2d}$ fixed. We obtain for Eq.$\eqref{hyst}$
\begin{equation}
\mathcal{A} \sim t_Q\cdot \Xi.
\end{equation}
where the constant $\Xi$ is given by the Eq.$\eqref{Xi}$. The energy spent by the system in a cycle has a rescaling:
\begin{equation}
\mathcal{W} = t_s^{-1}\cdot \mathcal{A}\sim (t_s)^{-1/2}\cdot \Xi,
\end{equation}
independently from the spatial dimensions considered.\\
Hysteresis phenomena across the FOT can be studied in O$(3)$ vector models. It has been shown that the off-equilibrium dynamics occurs in one of the $N-1$ transverse planes: we therefore expect that the hysteresis loop area and the magnetic work are almost similiar for any $N\geq 2$. Infact, the results above are in agreement with the numerical evidences \cite{VicOff}.\\

 \section{Asymptotic behaviors.}\label{SEC11}
 We investigate the first deviations from the equilibrium behavior in the correlation functions occuring at a time $|t|\sim t_Q$ before the transition. In terms of the rescaled time, the equilibrium has to be recovered in the asymptotic limit $t/t_Q \rightarrow -\infty$.\\
\subsection*{Case $T=T_c$}
Firstly, we consider the matching with the equilibrium for the continuous phase transition. By comparing the off-equilibrium scales $\eqref{KZ}$ with the instantaneous equilibrium correlation length $\xi(t,t_s)$ and time $\xi_t(t,t_s)$, it is possible to connect the off-equilibrium scaling regime with the equilibrium one \cite{CEGS-12}:
\begin{equation}
\frac{\xi(t,t_s)}{l_Q} \sim |t/t_Q|^{-\nu_g} \qquad \frac{\xi_t(t,t_s)}{t_Q} \sim |t/t_Q|^{-z\nu_g}.
\end{equation}
For the scaling function of the correlators in Eq.$\eqref{scalingoff}$, this means:
\[
\mathcal{G}^{(n)}(\overline{x}_1,\cdots,\overline{x}_n,\overline{t}) \sim \]
 \begin{equation}\label{match}
  |\overline{t}|^{\nu_g \alpha}\cdot \mathcal{G}^{(n)}_{\text{eq}} (\overline{x}_n\cdot |\overline{t}|^{\nu_g},\cdots ,\overline{x}_n \cdot |\overline{t}|^{\nu_g},\overline{t} \cdot |\overline{t}|^{z  \nu_g}).
 \end{equation}
 Approaching the equilibrium, the scaling functions of the correlators $\mathcal{G}^{(n)}$ present small fluctuations whose lifetime $\tau_o$ is of the order of the ratio between the two competing time-scales [see also \cite{VicOff}]:
 \begin{equation}
 \tau_o \sim \frac{\xi_t(t,t_s)}{t_Q} \sim |t/t_Q|^{-z\nu_g};
 \end{equation}
 We assume that these fluctuations are exponentially damped
  \[
 \mathcal{G}^{(n)}(\overline{x}_1, \cdots, \overline{x}_n ,\overline{t}) - \mathcal{G}^{(n)}_{\text{eq}} (\overline{x}_1, \cdot |\overline{t}| ^{\nu_g},\cdots ,\overline{x}_n \cdot |\overline{t}|^{\nu_g},\overline{t}  \cdot |\overline{t}|^{z  \nu_g}) \]
\begin{equation}  \overset{\overline{t}\rightarrow-\infty}{\sim}  
 \mathcal{G}^{(n)}_{\text{eq}} (\overline{x}_1, \cdot |\overline{t}| ^{\nu_g},\cdots ,\overline{x}_n \cdot |\overline{t}|^{\nu_g},\overline{t}  \cdot |\overline{t}|^{z  \nu_g}) \cdot K(\overline{t}) \cdot  e^{-C\frac{|\overline{t}|}{\tau_o}}.
 \end{equation}
 where $K$ is a regular function and $C$ is a positive constant. This ansatz has been numerically checked \cite{VicOff} for a magnetic field protocol $\eqref{protocol}$.\\
 Since in the limit of large $N$ the rescaled correlators $\mathcal{G}^{(n)}$ are uniquely determined by the value of the scaling function $\mathcal{M}^2$, we propose a general ansatz in terms of this function that is sufficient to reproduce asympotically the exponential damping of the off-equilibrium fluctuations in the observables. In the limit $t/t_Q \rightarrow -\infty$ we can write the scaling function
 \begin{equation}
 \mathcal{M}^2(t/t_Q) = \mathcal{M}^2_{\text{eq}} (\overline{t}) + \mathcal{M}^2_{\text{off}}(\overline{t})
 \end{equation}
 as an equilibrium term $\mathcal{M}^2_{\text{eq}}$ plus a very small off-equilibrium perturbation $\mathcal{M}^2_{\text{off}}\simeq 0$.\\
 At the equilibrium, we can relate the effective mass term of the system with the inverse of the instantaneous equilibrium correlation length. Therefore,
 \begin{equation}\begin{split}
 \mathcal{M}^2_{\text{eq}}(\overline{t}) \sim |\overline{t}|^{2\nu_g} + \text{high order}\\
\text{ corrections to the equilibrium scaling}
\end{split}\end{equation}
\\
By substituiting the leading equilibrium contribution $\mathcal{M}^2= \mathcal{M}^2_{\text{eq}}\simeq |\overline{t}|^{2\nu_g}$ in the expression of the scaling functions of the correlators $\eqref{magnscal}$, $\eqref{GTscal}$, the correct matching is recovered:
\begin{equation}
 \Theta(\overline{t}) \sim -|\overline{t}|^{d_{\phi}/d_h} \qquad  \chi_T(\overline{t})=\mathcal{G}_T(0,\overline{t}) \sim |\overline{t}|^{-2/d_h}
\end{equation}
in agreement with Eq.$\eqref{match}$ and with the standard definition of the equilibrium critical exponent $\delta=d_h/d_{\phi}$ and $\gamma=\nu_g(2-\eta)=2\nu_g$.\\
Let us write the leading off-equilibrium term assuming exponential approach to the equilibrium:
\begin{equation}
\mathcal{M}^2_{\text{off}}(\overline{t}) \sim |\overline{t}|^{2\nu_g}\cdot b\cdot |\overline{t}|^a \cdot e^{-c|\overline{t}|^{1+z\nu_g}} +\text{high orders}
\end{equation}
where $b$, $a$ and $c$ are constants. Thus, we can write 
\begin{equation}\label{d}\begin{split}
\mathcal{M}^2(\overline{t}) \overset{\overline{t} \rightarrow -\infty}{\sim} \{|\overline{t}|^{2\nu_g} +\cdots \} \cdot \Big( 1+\\
 b\cdot |t/t_Q|^a \cdot e^{-c|t/t_Q|^{1+z\nu_g}}+ O(e^{-2c|t/t_Q|^{1+z\nu_g}})\Big) \simeq \\
 \mathcal{M}^2_{\text{eq}}(\overline{t}) \cdot \Big( 1+ b\cdot |t/t_Q|^a \cdot e^{-c|t/t_Q|^{1+z\nu_g}}\Big).
\end{split}
\end{equation}
 Within this ansatz it is possible to compute the leading off-equilibrium corrections in the scaling functions of the magnetization and of the transverse susceptibility. The result, as expected, is an exponential approach:
 \begin{equation}\label{leadingmagn}
 \Theta(\overline{t}) \sim \Theta_{\text{eq}}(\overline{t})\cdot \Big(1+ K\cdot |\overline{t}|^a\cdot e^{-c|\overline{t}|^{1+z\nu_g}}\Big)
 \end{equation}
 where $K=-b/(c(1+z\nu_g)+1)$, and for the transverse susceptibility
 \begin{equation}\label{leadingsusc}
 \chi_{T}(\overline{t})=\mathcal{G}_T(0,\overline{t}) \sim\chi_{T,\text{eq}}(\overline{t}) \cdot \Big(1+K'\cdot |\overline{t}|^a \cdot e^{-c|\overline{t}|^{1+z\nu_g}}\Big)
 \end{equation}
 with $K'= -2b/(2+c(1+z\nu_g))$.\\
The  Eq.$\eqref{d}$ reproduces the exponential approach in the observables $\eqref{leadingmagn}$,  $\eqref{leadingsusc}$ and is consistent with the rescaled constraint equations $\eqref{a}$, $\eqref{b}$. Furthermore, starting with other types of assumptions for $\mathcal{M}^2_{\text{off}}$ such as a power law decay of the fluctuations, the same results does not follow and we lose the consistence.\\
 Note that these arguments apply also for a thermal quench $r(t,t_s)-r_c\approx -t/t_s$ leading to the same result for the susceptibility, if we use the appropriate value of the exponent $\nu_g=\nu= 1/(d-2)$ into $\eqref{leadingsusc}$. In general these arguments are valid for each type of protocol in the limit $t/t_Q\rightarrow -\infty$, for a system prepared at the equilibrium.\\
We also expect that the limit $t/t_Q \rightarrow +\infty$ exhibits the same behavior, maybe with different constants \cite{VicOff} in the case of a magnetic field protocol $\eqref{protocol}$. A thermal protocol undergoes coarsening \cite{Bray} after crossing the transition and presents a different matching for $t/t_Q \rightarrow +\infty$ \cite{CEGS-12}.\\	
Let us consider also the finite-size effects. In a finite geometry, a necessary condition to obtain equilibrium results is that $t_s \gg \tau$ i.e. $t_s\cdot \tau \rightarrow\infty$, where $\tau$ is the slowest time-scale of the system at the equilibrium given by $\tau \sim L^{z}$. Since $t_s\rightarrow \infty$ at fixed $\ell$ we have:
\begin{equation}
t_s\cdot L^{-z}=\ell^{z/e}\cdot L^{z(1-e)/z}
\end{equation}
This condition is satisfied only if $L\rightarrow \infty$. The previous relations implies that the matching occur at a time in which $\xi(t,t_s) < L$. Thus, the limit $\overline{t}\rightarrow \infty$ at fixed $\ell$ is expected to lead to the infinite-volume equilibrium behavior:
\begin{equation}\begin{split}
\mathcal{G}_{OO}(\overline{x}, \overline{t}, \ell) \sim |\overline{t}^a \cdot \ell^{-1/\nu_g}|^{2\nu_g\Delta}\cdot
  \mathcal{G}_{OO}^{\text{eq.}}(\overline{x}\cdot|\overline{t}|^{a\nu_g}, \overline{t}\cdot|\overline{t}|^{za\nu_g})
\end{split}
\end{equation}
The asymptotic behavior of a finite-size system matches the infinite-volume equilibrium scaling relations because occurs in a region with a finite correlation length [see also \cite{VicOff}].
\subsection*{Case $T<T_c$}
The same arguments can be applied also below the critical temperature in order to estimate the leading off-equilibrium corrections to the scaling behavior. The ansatz $\eqref{d}$ for the scaling function of the effective mass term now becomes:
\begin{equation}\label{dlow}
\mathcal{M}^2(\overline{t})\overset{\overline{t}\rightarrow -\infty}{\sim}  |\overline{t} /\sigma| \cdot \Big( 1+ b\cdot |\overline{t}|^a \cdot e^{-c|\overline{t}|^{1+z\nu_g}}\Big)
\end{equation}
 where $z\nu_g=1$ and $a$, $b$, $c$ are constants. The equilibrium contribution $\mathcal{M}^2_{\text{eq}}(\overline{t})\simeq |\overline{t}/\sigma|$ is given by the scaling limit of Eq.$\eqref{a}$, where we have considered the magnetization as a constant. Using the Eq.$\eqref{dlow}$, one can compute the leading off-equilibrium corrections in the scaling behavior of the correlators. For instance, the first deviation in the scaling function of the magnetization $\eqref{magn}$ are given by:
 \begin{equation}
\frac{\Theta(\overline{t})}{|\sigma|} \overset{\overline{t}\rightarrow -\infty}{\sim} -1\Big(1+ K'' \cdot |\overline{t}|^a \cdot e^{-c|\overline{t}|^2}\Big).
 \end{equation}
 with $K''=(-b/(1+2c|\sigma|)$. Even for $T<T_c$, the exponential approach to the equilibrium has been numerically verified in \cite{VicOff}. The exponential approach to the equilibrium is also consistent with the constraint Eq.s $\eqref{a}$ and $\eqref{b}$. Furthermore, the analysis of the asymptotic matching with the equilibrium is not modified if we consider a system with a finite size $L$, as in the case $T=T_c$.\\
 
 \hrulefill\\
 
 We finally note that the ansatz of exponential approach to the equilibrium is sufficient to obtain a finite rescaled hysteresis loop area $\Xi <\infty$ for both the cases $T=T_c$ and $T<T_c$.

 \section{Conclusions.}\label{SEC12}
 
We study the slow passage through the critical point of a statistical system in the presence of a time-dependent magnetic field $h(t,t_s)\approx t/t_s$, where $t_s$ is a time scale, focusing on a spin system with O$(N)$ symmetry. This model shows a continuous phase transition occuring at the critical temperature $T=T_c$ and at zero magnetic field $h=0$. Very close to the critical point $h(t,t_s)\simeq 0$, the system goes out of the equilibrium because it develops large scale modes which cannot adapt themselves to the variations of the external parameters, even in the limit of slow passage $t_s\rightarrow\infty$. The dynamics of the system shows universal scaling behaviors, which are controlled by the time $t$ and the time scale $t_s$. In this regime the time dependence of the correlations can be expressed in terms of universal scaling functions that depend on the scaling variables $t/(t_s)^e=t/t_Q$ and $x/(t_s)^{e/z}=x/l_Q$, where $z$ is the dynamical critical exponent and $0<e<1$ is a universal exponent depending on the static universality class of the model, on the type of dynamics and on the behavior of the specific protocol near the transition. The magnetic field protocol was numerically studied \cite{VicOff}, we provide to analytical computations in the limit of large $N$  and we demonstrate the existence of a non-trivial rescaling very close to the critical point. We check that these relations for the O$(N)$ vector model at large $N$ are satisfied only if the rescaling is made with $l_Q$ and $t_Q$. The prediction for the scaling relations at large $N$ are in agreement with the numerical result for the case $N=3$ \cite{VicOff}. The large $N$ limit does not modify the qualitatively off-equilibrium behavior of the system and the scaling relations apply for finite $N$ with appropriate exponents.\\ 
We also show how to construct an analogue of the KZ approach for describing the dynamics below the critical temperature, where the O$(N)$ vector model undergoes a first-order phase transition at $h=0$. For a magnetic field protocol $h(t,t_s)\approx t/t_s$ at $T<T_c$, we derive the off-equilibrium scaling relations and the scaling functions for the correlators, pointing out the analogies with the case $T=T_c$. Below the critical temperature, a constraint equation of the O$(N)$ vector model at large $N$ predicts that the magnetization of the system behaves as a rigid spin under the effects of a time-dependent magnetic fields and makes a slowly rotation in the off-equilibrium region.\\
For both the cases $T\leq T_c$, we perform the study of the first deviations from the equilibrium scaling behavior occuring at a time $|t|\sim t_Q$ before the transition. We expect that the fluctuations over the equilibrium background in the correlation functions, decay exponentially with a lifetime of the order of the ratio between the equilibrium relaxation time and the off-equilibrium time scale. The same behavior is shown by the system also after the transition if it approaches again the equilibrium. In particular, we demonstrate that is sufficient to formulate an ansatz in terms of the scaling function of the effective mass term  $\mathcal{M}^2$ of the O$(N)$ vector model at large $N$ to have an exponential decay in the correlation functions. We verify the consistence of this ansatz in both the cases.\\
We also investigate the off-equilibrium behavior arising when the system is coupled to a magnetic field which varies in time from $t_i<0$ to $t_f>0$ and then back from $t_f$ to $t_i$. The system presents hysteresis phenomena related to the off-equilibrium. The area of the hysteresis obeys to a scaling relation and can be easly connected to the magnetic work done by the system over a round-trip protocol. For the case $T=T_c$, we obtained a scaling relation for the magnetic work in three spatial dimensions:
\[
\mathcal{W}\sim (t_s)^{-2/3}\cdot \Xi,\]
where $\Xi$ is an amplitude constant which is finite under the assumption of exponential decay abovementioned. Extending this result to the case $N=3$ using the critical exponents of the Heisenberg universality class, we found that the magnetic work scales as
\[
\mathcal{W}^{(N=3)} \sim (t_s)^{-0.66}\cdot \Xi^{(N=3)}\]
which is in agreement with the numerical results \cite{VicOff}. Hysteresis phenomena are shown by the system also in the ordered phase when we consider a round-trip protocol. The scaling relation of the magnetic work is:
\[
\mathcal{W}\sim (t_s)^{-1/2}\cdot \Xi.\]
where $\Xi$ is a finite constant under the assumption of exponential damping. \\
\\
\textit{Acknowledgments.}\\
I wish to thank Ettore Vicari for the useful discussions and for the support in the development of this work. Without his help, this paper would never has been written.

\end{document}